\documentclass[prl,twocolumn]{revtex4}

 \usepackage[english]{babel}
 \usepackage{epsfig,amssymb,amsmath}
 \usepackage{color}
 \usepackage{pstricks}

\begin{document}

\title{Single-spin polaron memory effect}

\author{Dmitry\,A.~Ryndyk, Pino~D'Amico, and Klaus~Richter}

\affiliation{Institute for Theoretical Physics, University of Regensburg,
D-93040 Regensburg, Germany}

\date{\today}

\begin{abstract}
The  single-spin memory effect is considered within a minimal polaron model
describing a single-level quantum dot interacting with a vibron and weakly
coupled to ferromagnetic leads. We show that in the case of strong
electron-vibron and Coulomb interactions the rate of spontaneous quantum
switching between two spin states is suppressed at zero bias voltage, but can
be tuned through a wide range of finite switching timescales upon changing the
bias. We further find that such junctions exhibit hysteretic behavior enabling
controlled switching of a spin state. Spin lifetime, current and spin
polarization are calculated as a function of the bias voltage by the master
equation method. We also propose to use a third tunneling contact to control
and readout the spin state.
\end{abstract}

\maketitle

One of the most promising directions in the fields of molecular electronics and
spintronics is the experimental and theoretical investigation of spin
manipulation in quantum dots and single molecules. In particular, new methods
have been recently developed to investigate spin states of single atoms and
molecules using spin-polarized scanning tunneling
spectroscopy~\cite{Meier08science,Iacovita08prl}. Motivated by such
achievements the promising question arises whether a single-spin memory effect
(including bistability and controlled switching between spin states) is
possible.

One of the ways for single spin manipulation is based on the interplay of
charge, spin, and vibron degrees of freedom in molecular junctions. In various
experiments the signatures of the electron-vibron (e-v) interaction have been
observed in atomic scale
structures~\cite{Qiu04prl,Wu04prl,Repp05prl,Repp05prl2}. In the case of strong
e-v interaction the formation of a local polaron can lead to a charge-memory
effect, which was first predicted long time ago~\cite{Hewson79jphysc}, and has
been recently reconsidered in more
detail~\cite{Alexandrov03prb,Galperin05nanolett,Mitra05prl,Mozyrsky06prb,Ryndyk08prb,DAmico08njp}.
Neutral and charged (polaron) states correspond to different local minima of an
effective energy surface and are metastable if the e-v interaction is strong
enough. By applying an external voltage, one can change the charge state of
this bistable system, an effect that is accompanied by hysteretic
charge-voltage and current-voltage curves. A similar memory effect was found in
recent STM experiments~\cite{Repp04science,Olsson07prl} as a multistability of
neutral and charged states of single metallic atoms coupled to a metallic
substrate through a thin insulating ionic film; the corresponding polaron model
was discussed in~\cite{Ryndyk08prb,DAmico08njp}.

\begin{figure}[b]
\begin{center}
\epsfxsize=0.75\hsize \vskip 0.3cm
\epsfbox{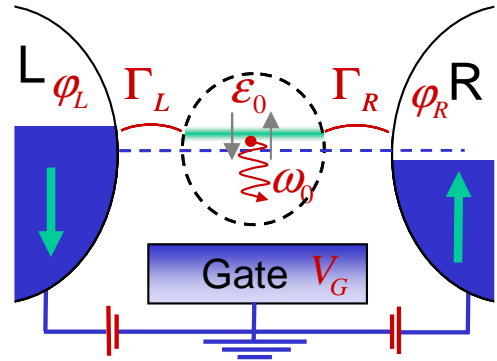}
\caption{(Color online) Schematic picture of the considered system: a gated
single-level quantum dot interacting with a vibron and coupled to ferromagnetic leads.}
\label{fig1}
\end{center}
\end{figure}

In this Letter we propose an approach to observe a single-spin memory effect by
combining the polaron memory mechanism and the spin-dependent tunneling. To
this end we consider a single-level and single-vibron quantum system between
magnetic leads (Fig.\,\ref{fig1}). We study the case of a symmetric junction
with anti-parallel magnetizations of left and right leads, besides the third
electrode can be used as a gate or to probe the spin state. The problem can be
solved with well controlled approximation in the limit of weak coupling to the
leads, where the master equation for sequential tunneling can be used. Thus we
focus our major discussion on this limit.

The Hamiltonian of the single-level polaron (Anderson-Holstein) model is
\begin{align}\label{H-SL}
& \hat H =\sum_\sigma\tilde\epsilon_\sigma d_\sigma^{\dag}d_\sigma+\omega_0a^{\dag}a+
  \lambda\left(a^{\dag}+a\right)\hat{n}+U\hat{n}_\uparrow\hat{n}_\downarrow \nonumber \\
+ & \sum_{ik\sigma}\left[(\epsilon_{ik\sigma}+e\varphi_i)
  c^{\dag}_{ik\sigma}c_{ik\sigma}+\left(V_{ik\sigma}c^{\dag}_{ik\sigma}d_\sigma+h.c.\right)\right].
\end{align}
Here the first line describes the free electron states with energies
$\tilde\epsilon_\sigma$, the free vibron of frequency $\omega_0$, the
electron-vibron and Coulomb interactions with coupling strength $\lambda$ and
$U$, respectively; $\sigma$ is the spin index and
$\hat{n}_\sigma=d_\sigma^{\dag}d_\sigma$,
$\hat{n}=\hat{n}_\uparrow+\hat{n}_\downarrow$. The other terms are the
Hamiltonian of the leads and the tunneling coupling ($i=L,R$ is the lead index,
$k$ labels the electronic states). The bias voltage $V$ is introduced through
the left and right electrical potentials, $V=\varphi_L-\varphi_R$. The energy
$\tilde\epsilon_\sigma=\epsilon_\sigma+e\varphi_0$ includes the bare level
energies ($\epsilon_\uparrow=\epsilon_\downarrow=\epsilon_0$ below) and the
electrical potential $\varphi_0$ describing the shift of the central level by
the gate voltage $V_G$ and the bias voltage drop between the left and right
lead: $\varphi_0=\varphi_R+\eta(\varphi_L-\varphi_R)+\alpha V_G$, where $0 <
\eta < 1$ describes the symmetry of the voltage drop across the junction,
$\eta=0.5$ stands for the symmetric case considered below.

The coupling to the leads is characterized by the level-width function
\begin{equation}
\Gamma_{i\sigma}(\epsilon)=2\pi\sum_{k}|V_{ik\sigma}|^2\delta(\epsilon-\epsilon_{ik\sigma}).
\end{equation}
In the wide-band limit considered below, the spin-dependent densities of states
in the leads and the tunneling matrix elements are assumed to be
energy-independent, so that $\Gamma_{L\sigma}$ and $\Gamma_{R\sigma}$ are
constants. The full level broadening is given by the sum
\mbox{$\Gamma_\sigma=\Gamma_{L\sigma}+\Gamma_{R\sigma}$}. Below we consider a
symmetric junction with antiparallel magnetization of the leads and use the
notation $\Gamma_{L\downarrow}=\Gamma_{R\uparrow}=\Gamma$ for majority spins
and $\Gamma_{L\uparrow}=\Gamma_{R\downarrow}=\kappa\Gamma$ for minority spins,
$\kappa\ll 1$.

The spin effects addressed are particularly pronounced in the limit
$U\rightarrow\infty$, i.e. we neglect the doubly occupied state, so that only
three states in the charge sector should be considered: neutral $|0\rangle$,
charged spin-up $|\!\!\uparrow\rangle$ and charged spin-down
$|\!\!\downarrow\rangle$. Using the polaron
(Lang-Firsov)~\cite{Lang63jetp,Hewson74jjap,Mahan90book} canonical
transformation, the eigenstates of the {\em isolated system} ($\Gamma=0$) are
\begin{equation}\label{V-P-ES2-Spin-1}
  |\psi_{0q}\rangle=
  \frac{(a^\dag)^{q}}{\sqrt{q!}}|0\rangle,
\end{equation}
\begin{equation}\label{V-P-ES2-Spin-2}
  |\psi_{\sigma q}\rangle=
  e^{-\frac{\lambda}{\omega_0}\left(a^\dag-a\right)d_\sigma^\dag d_\sigma}
  d_\sigma^\dag\frac{(a^\dag)^{q}}{\sqrt{q!}}|0\rangle,
\end{equation}
with the eigenenergies
\begin{equation}\label{V-P-EE-Spin}
  E_{0q}=\omega_0q,\ \
  E_{\sigma q}=\tilde\epsilon'_\sigma+\omega_0q, \ \ \tilde\epsilon'_\sigma=\tilde\epsilon_\sigma-\frac{\lambda^2}{\omega_0},
\end{equation}
where the quantum number $q$ characterizes vibronic eigenstates, which are
superpositions of states with different number of bare vibrons.

Taking into account all possible single-electron tunneling processes for both
leads, we obtain the incoming and outgoing tunneling rates
\begin{align}\label{Gin-Spin-1}
  \Gamma^{\sigma 0}_{qq'} &=\sum_{i=L,R}\Gamma^{\sigma 0}_{iqq'}=
  \sum_{i=L,R}\Gamma_{i\sigma} \left|M_{qq'}\right|^2
  f_i^0(E_{\sigma q}-E_{0q'}) \nonumber \\
  & =\sum_{i=L,R}\Gamma_{i\sigma} \left|M_{qq'}\right|^2 f_i^0(\tilde\epsilon'_\sigma+\omega_0(q-q')), \\
%
  \Gamma^{0\sigma}_{qq'} & =\sum_{i=L,R}\Gamma^{0\sigma}_{iqq'}=
  \sum_{i=L,R}\Gamma_{i\sigma} \left|M_{qq'}\right|^2 \left(1-f_i^0(E_{\sigma q'}-E_{0q})\right)\nonumber \\
  & =\sum_{i=L,R}\Gamma_{i\sigma} \left|M_{qq'}\right|^2 \left(1-f_i^0(\tilde\epsilon'_\sigma-\omega_0(q-q'))\right). \label{Gin-Spin-2}
\end{align}
Here $f_i^0(\epsilon)$ is the equilibrium Fermi function in the lead shifted by
the external potential, $f_i^0(\epsilon)=f^0(\epsilon-e\varphi_i)$, and
$M_{qq'}$ is the Franck-Condon matrix element that can be calculated
analytically (see
Refs.\,\cite{Braig03prb,Mitra04prb,Koch05prl,Koch06prb,Koch06prb2} for details
of the master equation method and calculation of the tunneling rates). The
incoming rate $\Gamma^{\sigma 0}_{qq'}$ describes tunneling of one electron
with spin $\sigma$ from the lead to the dot changing the state of the dot from
$|0q'\rangle$ to $|\sigma q\rangle$. The outgoing rate $\Gamma^{0\sigma}_{qq'}$
corresponds to the transition from $|\sigma q'\rangle$ to $|0q\rangle$.

\begin{figure}[t]
\begin{center}
\epsfxsize=0.8\hsize \vskip 0.3cm
\epsfbox{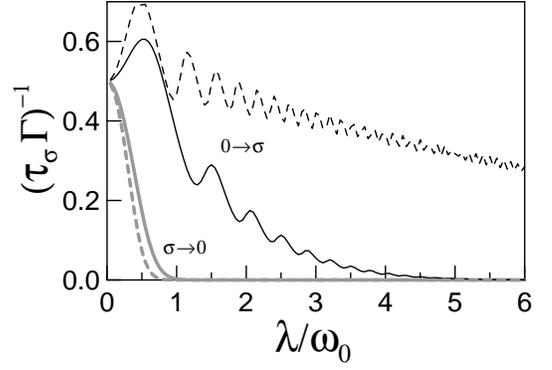}
\caption{Inverse lifetime $\gamma^{\sigma0}/\Gamma$ of the neutral state (thin solid line) and
the inverse spin lifetime $(\tau_\sigma\Gamma)^{-1}$ (thick gray solid line)
as a function of the scaled electron-vibron coupling $\lambda/\omega_0$ at $\epsilon_0=\lambda^2/2\omega_0$ and at
$\epsilon_0=0.1\lambda^2/\omega_0$ (corresponding dashed lines), $T=0.1\omega_0$. }
\label{fig2}
\end{center}
\end{figure}

In the sequential tunneling regime the master equation for the probability
$P^n_q(t)$, $n=0,\uparrow,\downarrow$ to find the system in one of the polaron
eigenstates (\ref{V-P-ES2-Spin-1}), (\ref{V-P-ES2-Spin-2}) can be written
as~\cite{Braig03prb,Mitra04prb,Koch05prl,Koch06prb,Koch06prb2}
\begin{equation}\label{V-ST-ME}
  \frac{dP^n_q}{dt}=\sum_{n'q'}\Gamma^{nn'}_{qq'}P^{n'}_{q'}-
  \sum_{n'q'}\Gamma^{n'n}_{q'q}P^n_q+I^V[P].
\end{equation}
Here the first term describes the tunneling transition {\em into the state}
$|nq \rangle$ and the second term the transition {\em out of the state}
$|nq\rangle$. $I^V[P]$ is the vibron scattering integral describing the
relaxation of the vibrons to the thermal equilibrium.

Finally, the average charge and the spin polarization are
\begin{equation}
  Q=e\sum_{q}\left(P^\uparrow_q+P^\downarrow_q\right),\ \  S=\sum_{q}\left(P^\uparrow_q-P^\downarrow_q\right),
\end{equation}
respectively, and the average current (from the left or right lead) reads
\begin{equation}
J_{i=L,R}=e\sum_{\sigma qq'}
\left(\Gamma^{\sigma 0}_{iqq'}P^0_{q'}-\Gamma^{0\sigma}_{iqq'}P^\sigma_{q'}\right).
\end{equation}

To proceed further, we calculate the characteristic lifetimes of the neutral,
spin-up, and spin-down ground states ($q=0$). We define the switching rates
$\gamma^{\sigma 0}$ from the neutral to the charged state with spin $\sigma$
and vice-versa as the sum of the rates of all possible processes which change
these states
\begin{equation}\label{gamma_sigma_0}
\gamma^{\sigma 0}=\sum_{q}\Gamma^{\sigma 0}_{q0},\ \
\gamma^{0\sigma}=\sum_{q}\Gamma^{0\sigma}_{q0}.
\end{equation}

In the sequential tunneling approximation the spin lifetime $\tau_\sigma$  is
determined by the lifetime of the charged state, from (\ref{Gin-Spin-2}),
(\ref{gamma_sigma_0}). It reads (assuming that the Fermi energy in the leads is
zero, see details in~\cite{Ryndyk08prb}, $g=(\lambda/\omega_0)^2$)
\begin{equation}
  \tau^{-1}_\sigma=\gamma^{0\sigma}=(1+\kappa)\Gamma\sum_{q}\frac{e^{-g}g^{q}}{q!}
  f^0\left(-\tilde\epsilon'_\sigma+\omega_0q\right).
\end{equation}
At large $g$ the sequential tunneling rates are exponentially suppressed and
the cotunneling contribution to $\tau^{-1}_\sigma$ becomes dominant. It can be
estimated as \cite{Koch06prb2}
\begin{equation}
  \tau^{-1 (ct)}_\sigma\approx\frac{\kappa\Gamma^2T\omega_0^2}{\lambda^4}.
\end{equation}
Although the cotunneling contribution is not suppressed exponentially by
Franck-Condon blockade, it is of the second order in the tunneling coupling and
suppressed additionally by the small polarization parameter $\kappa$ and large
$\lambda$. At typical parameters, considered in this Letter, the cotunneling
contribution can be neglected, but it can be essential at larger tunneling
couplings and larger temperatures.

\begin{figure}[t]
\begin{center}
\epsfxsize=0.8\hsize \vskip 0.3cm
\epsfbox{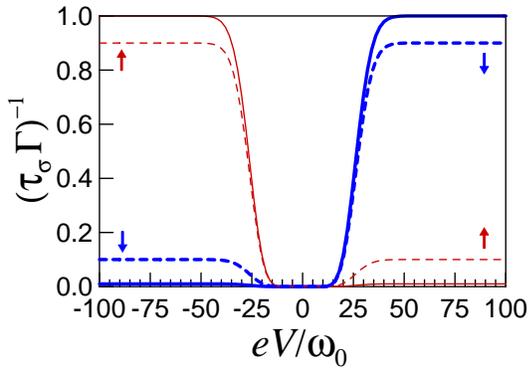}
\caption{(Color online) Inverse spin lifetime as a function of normalized bias voltage
$eV/\omega_0$ at $\lambda/\omega_0=3$, $\kappa=0.01$, $\epsilon_0=\lambda^2/2\omega_0$
for the spin-up state (thin red solid line) and the spin-down state (thick blue solid line)
and the same for a less polarized junction ($\kappa=0.1$, dashed lines).}
\label{fig3}
\end{center}
\end{figure}

The dependence of $\tau^{-1}_\sigma$ and $\gamma^{\sigma0}$ on the scaled
electron-vibron interaction constant $\sqrt{g}=\lambda / \omega_0$ is shown in
Fig.~\ref{fig2}. For large values of $\lambda$ the tunneling from the neutral
state to the charged state and vice versa is suppressed compared to the bare
tunneling rate $\Gamma$. Hence all states are (meta)stable at low temperatures
and zero voltage. Moreover, the lifetime of the charged states can be much
larger than that of the neutral state.

Next we address the other important question, whether fast switching between
the two spin states is possible. To this end we consider what happens, if one
sweeps the voltage with different velocities, $\tau_{exp}$ is the
characteristic time of the voltage change. At this point an assumption about
the relaxation time $\tau_V$ of the vibrons without change of the charge state
is due. We assume that the relaxation is fast, $\tau_V\ll
\tau_\sigma,\tau_{exp}$, so that after an electron tunneling event the system
relaxes rapidly into the vibronic ground state $|\sigma 0\rangle$ or
$|00\rangle$. In this case the probabilities $P^\sigma=\sum_q P^\sigma_q$ of
the charged and $P^0=\sum_q P^0_q$ of the neutral state are determined from the
equations
\begin{align}
& \frac{dP^0}{dt}=\sum_\sigma\left(\gamma^{0\sigma}P^\sigma-\gamma^{\sigma 0}P^0\right),  \\
& \frac{dP^\sigma}{dt}=\gamma^{\sigma 0}P^0-\gamma^{0\sigma}P^\sigma,
\end{align}
where the switching rates $\gamma^{\sigma 0}$, $\gamma^{0\sigma}$ at finite
voltage are calculated from
Eqs.\,(\ref{Gin-Spin-1},\ref{Gin-Spin-2},\ref{gamma_sigma_0}):
\begin{align}\nonumber
  \gamma^{\sigma 0}=\sum_{q}\frac{e^{-g}g^{q}}{q!} & \left[
  \Gamma_{L\sigma}f^0\left(\tilde\epsilon'_\sigma+\omega_0q-(1-\eta)eV\right)\right. \\ \label{V-ST-T4}
  & +\left.\Gamma_{R\sigma}f^0\left(\tilde\epsilon'_\sigma+\omega_0q+\eta eV\right)
  \right], \\
%
\nonumber
  \gamma^{0\sigma}=\sum_{q}\frac{e^{-g}g^{q}}{q!} & \left[
  \Gamma_{L\sigma}f^0\left(-\tilde\epsilon'_\sigma+\omega_0q+(1-\eta)eV\right)\right. \\ \label{V-ST-T5}
  & +\left.\Gamma_{R\sigma}f^0\left(-\tilde\epsilon'_\sigma+\omega_0q-\eta eV\right)
  \right].
\end{align}

\begin{figure}[t]
\begin{center}
\epsfxsize=0.8\hsize \vskip 0.3cm
\epsfbox{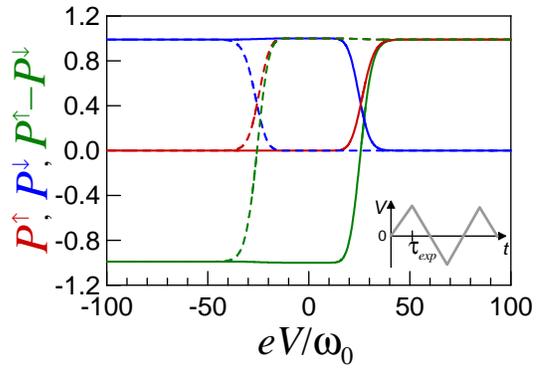}
\caption{(Color online) Populations of the spin-up state (red), spin-down state (blue), and
spin polarization (green) as a function of normalized voltage $eV/\omega_0$ at $\lambda/\omega_0=3$ and
$\epsilon_0=\lambda^2/2\omega_0$, the solid (dashed) for increasing (decreasing)voltage.
Inset: sketch of voltage time-dependence.}
\label{fig4}
\end{center}
\end{figure}

\begin{figure}[t]
\begin{center}
\epsfxsize=0.8\hsize \vskip 0.3cm
\epsfbox{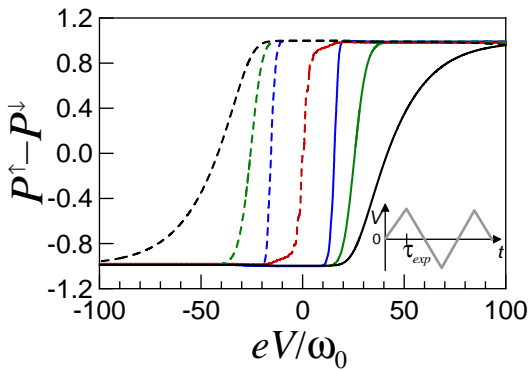}
\caption{(Color online) Spin polarization as a function of normalized voltage
$eV/\omega_0$ at $\lambda/\omega_0=3$ and
$\epsilon_0=\lambda^2/2\omega_0$ for three different sweep velocities relative to
that in Fig.\ref{fig4} (here shown by green):
faster (black), slower (blue) and in the adiabatic limit (red dashed line).}
\label{fig5}
\end{center}
\end{figure}

The voltage dependence of the inverse spin lifetime is depicted in
Fig.~\ref{fig3}. If the voltage is large enough, the Franck-Colomb blockade is
overcome and the system is switched into spin-up (spin-down) state at positive
(negative) voltage. If the bias voltage is swept fast enough, i.e. faster than
the spin lifetime at zero voltage, $\tau_{exp}\ll\tau_\sigma(0)$, both spin
states can be considered as stable at zero voltage and hysteresis takes place.
This is shown in Fig.~\ref{fig4} where the solid (dashed) lines mark the spin
population for increasing (decreasing) bias voltage. In the opposite
(adiabatic) limit the voltage change is so slow that the system relaxes into
the equilibrium state, and the population-voltage curve is single-valued
(Fig.~\ref{fig5}).

\begin{figure}[b]
\begin{center}
\epsfxsize=0.77\hsize \vskip 0.3cm
\epsfbox{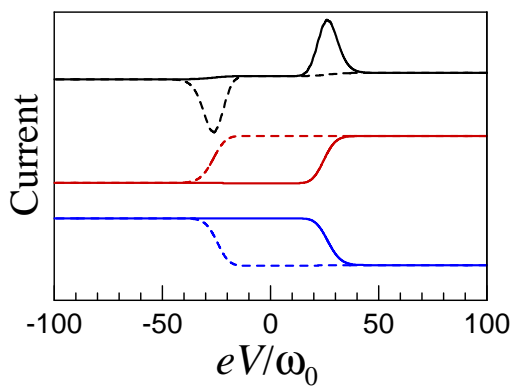}
\caption{(Color online) Bias current (top black) and test current for spin-up
(middle red) and spin-down (bottom blue) test electrode
magnetization as a function of normalized voltage $eV/\omega_0$;
all other parameters are the same as in Fig.\,\ref{fig4}. The curves are shifted along vertical axes
and the magnitude of the test current is much smaller than of the bias current.}
\label{fig6}
\end{center}
\end{figure}

Finally, we study the signatures of the spin polarization in the charge current
which is most easily accessible to experiments. In Fig.~\ref{fig6} we show the
bias current and the test current to the additional ferromagnetic electrode,
very weakly coupled to the system, so that it does not perturb the state. At
large negative voltage applied to the electrode the current is sensitive to the
orientation of the magnetization in the test electrode, thus the spin state can
be controlled during the experiment. Also such a small current can be used to
readout the memory element.

In conclusion, we considered a {\em single-spin memory effect} and switching
phenomena in the framework of a single-level quantum dot polaron model, taking
into account non-stationary effects, in particular the interplay between the
timescales of voltage sweeping and the quantum switching rates of meta-stable
states. We showed that the bistability arises when the quantum switching
between two spin states is suppressed due to the Franck-Condon blockade.
Controlled switching of the spin can be achieved by applying finite voltage
pulses.

This work was funded by the Deutsche Forschungsgemeinschaft within the Priority
Program SPP 1243 and Collaborative Research Center SFB 689.

\bibliography{}

\begin{thebibliography}{23}
\expandafter\ifx\csname natexlab\endcsname\relax\def\natexlab#1{#1}\fi
\expandafter\ifx\csname bibnamefont\endcsname\relax
  \def\bibnamefont#1{#1}\fi
\expandafter\ifx\csname bibfnamefont\endcsname\relax
  \def\bibfnamefont#1{#1}\fi
\expandafter\ifx\csname citenamefont\endcsname\relax
  \def\citenamefont#1{#1}\fi
\expandafter\ifx\csname url\endcsname\relax
  \def\url#1{\texttt{#1}}\fi
\expandafter\ifx\csname urlprefix\endcsname\relax\def\urlprefix{URL }\fi
\providecommand{\bibinfo}[2]{#2}
\providecommand{\eprint}[2][]{\url{#2}}

\bibitem[{\citenamefont{Iacovita et~al.}(2008)\citenamefont{Iacovita, Rastei,
  Heinrich, Brumme, Kortus, Limot, and Bucher}}]{Iacovita08prl}
\bibinfo{author}{\bibfnamefont{C.}~\bibnamefont{Iacovita}},
  \bibinfo{author}{\bibfnamefont{M.~V.} \bibnamefont{Rastei}},
  \bibinfo{author}{\bibfnamefont{B.~W.} \bibnamefont{Heinrich}},
  \bibinfo{author}{\bibfnamefont{T.}~\bibnamefont{Brumme}},
  \bibinfo{author}{\bibfnamefont{J.}~\bibnamefont{Kortus}},
  \bibinfo{author}{\bibfnamefont{L.}~\bibnamefont{Limot}}, \bibnamefont{and}
  \bibinfo{author}{\bibfnamefont{J.~P.} \bibnamefont{Bucher}},
  \bibinfo{journal}{Phys. Rev. Lett.} \textbf{\bibinfo{volume}{101}},
  \bibinfo{pages}{116602} (\bibinfo{year}{2008}).

\bibitem[{\citenamefont{Meier et~al.}(2008)\citenamefont{Meier, Zhou, Wiebe,
  and Wiesendanger}}]{Meier08science}
\bibinfo{author}{\bibfnamefont{F.}~\bibnamefont{Meier}},
  \bibinfo{author}{\bibfnamefont{L.}~\bibnamefont{Zhou}},
  \bibinfo{author}{\bibfnamefont{J.}~\bibnamefont{Wiebe}}, \bibnamefont{and}
  \bibinfo{author}{\bibfnamefont{R.}~\bibnamefont{Wiesendanger}},
  \bibinfo{journal}{Science} \textbf{\bibinfo{volume}{320}},
  \bibinfo{pages}{82} (\bibinfo{year}{2008}).

\bibitem[{\citenamefont{Qiu et~al.}(2004)\citenamefont{Qiu, Nazin, and
  Ho}}]{Qiu04prl}
\bibinfo{author}{\bibfnamefont{X.~H.} \bibnamefont{Qiu}},
  \bibinfo{author}{\bibfnamefont{G.~V.} \bibnamefont{Nazin}}, \bibnamefont{and}
  \bibinfo{author}{\bibfnamefont{W.}~\bibnamefont{Ho}}, \bibinfo{journal}{Phys.
  Rev. Lett.} \textbf{\bibinfo{volume}{92}}, \bibinfo{pages}{206102}
  (\bibinfo{year}{2004}).

\bibitem[{\citenamefont{Wu et~al.}(2004)\citenamefont{Wu, Nazin, Chen, Qiu, and
  Ho}}]{Wu04prl}
\bibinfo{author}{\bibfnamefont{S.~W.} \bibnamefont{Wu}},
  \bibinfo{author}{\bibfnamefont{G.~V.} \bibnamefont{Nazin}},
  \bibinfo{author}{\bibfnamefont{X.}~\bibnamefont{Chen}},
  \bibinfo{author}{\bibfnamefont{X.~H.} \bibnamefont{Qiu}}, \bibnamefont{and}
  \bibinfo{author}{\bibfnamefont{W.}~\bibnamefont{Ho}}, \bibinfo{journal}{Phys.
  Rev. Lett.} \textbf{\bibinfo{volume}{93}}, \bibinfo{pages}{236802}
  (\bibinfo{year}{2004}).

\bibitem[{\citenamefont{Repp et~al.}(2005{\natexlab{a}})\citenamefont{Repp,
  Meyer, Stojkovi\'{c}, Gourdon, and Joachim}}]{Repp05prl}
\bibinfo{author}{\bibfnamefont{J.}~\bibnamefont{Repp}},
  \bibinfo{author}{\bibfnamefont{G.}~\bibnamefont{Meyer}},
  \bibinfo{author}{\bibfnamefont{S.~M.} \bibnamefont{Stojkovi\'{c}}},
  \bibinfo{author}{\bibfnamefont{A.}~\bibnamefont{Gourdon}}, \bibnamefont{and}
  \bibinfo{author}{\bibfnamefont{C.}~\bibnamefont{Joachim}},
  \bibinfo{journal}{Phys. Rev. Lett.} \textbf{\bibinfo{volume}{94}},
  \bibinfo{pages}{026803} (\bibinfo{year}{2005}{\natexlab{a}}).

\bibitem[{\citenamefont{Repp et~al.}(2005{\natexlab{b}})\citenamefont{Repp,
  Meyer, Paavilainen, Olsson, and Persson}}]{Repp05prl2}
\bibinfo{author}{\bibfnamefont{J.}~\bibnamefont{Repp}},
  \bibinfo{author}{\bibfnamefont{G.}~\bibnamefont{Meyer}},
  \bibinfo{author}{\bibfnamefont{S.}~\bibnamefont{Paavilainen}},
  \bibinfo{author}{\bibfnamefont{F.~E.} \bibnamefont{Olsson}},
  \bibnamefont{and} \bibinfo{author}{\bibfnamefont{M.}~\bibnamefont{Persson}},
  \bibinfo{journal}{Phys. Rev. Lett.} \textbf{\bibinfo{volume}{95}},
  \bibinfo{pages}{225503} (\bibinfo{year}{2005}{\natexlab{b}}).

\bibitem[{\citenamefont{Hewson and Newns}(1979)}]{Hewson79jphysc}
\bibinfo{author}{\bibfnamefont{A.~C.} \bibnamefont{Hewson}} \bibnamefont{and}
  \bibinfo{author}{\bibfnamefont{D.~M.} \bibnamefont{Newns}},
  \bibinfo{journal}{J. Phys. C: Solid State Phys.}
  \textbf{\bibinfo{volume}{12}}, \bibinfo{pages}{1665} (\bibinfo{year}{1979}).

\bibitem[{\citenamefont{Alexandrov and Bratkovsky}(2003)}]{Alexandrov03prb}
\bibinfo{author}{\bibfnamefont{A.~S.} \bibnamefont{Alexandrov}}
  \bibnamefont{and} \bibinfo{author}{\bibfnamefont{A.~M.}
  \bibnamefont{Bratkovsky}}, \bibinfo{journal}{Phys. Rev. B}
  \textbf{\bibinfo{volume}{67}}, \bibinfo{pages}{235312}
  (\bibinfo{year}{2003}).

\bibitem[{\citenamefont{Galperin et~al.}(2005)\citenamefont{Galperin, Ratner,
  and Nitzan}}]{Galperin05nanolett}
\bibinfo{author}{\bibfnamefont{M.}~\bibnamefont{Galperin}},
  \bibinfo{author}{\bibfnamefont{M.~A.} \bibnamefont{Ratner}},
  \bibnamefont{and} \bibinfo{author}{\bibfnamefont{A.}~\bibnamefont{Nitzan}},
  \bibinfo{journal}{Nano Lett.} \textbf{\bibinfo{volume}{5}},
  \bibinfo{pages}{125} (\bibinfo{year}{2005}).

\bibitem[{\citenamefont{Mitra et~al.}(2005)\citenamefont{Mitra, Aleiner, and
  Millis}}]{Mitra05prl}
\bibinfo{author}{\bibfnamefont{A.}~\bibnamefont{Mitra}},
  \bibinfo{author}{\bibfnamefont{I.}~\bibnamefont{Aleiner}}, \bibnamefont{and}
  \bibinfo{author}{\bibfnamefont{A.~J.} \bibnamefont{Millis}},
  \bibinfo{journal}{Phys. Rev. Lett.} \textbf{\bibinfo{volume}{94}},
  \bibinfo{pages}{076404} (\bibinfo{year}{2005}).

\bibitem[{\citenamefont{Mozyrsky et~al.}(2006)\citenamefont{Mozyrsky, Hastings,
  and Martin}}]{Mozyrsky06prb}
\bibinfo{author}{\bibfnamefont{D.}~\bibnamefont{Mozyrsky}},
  \bibinfo{author}{\bibfnamefont{M.~B.} \bibnamefont{Hastings}},
  \bibnamefont{and} \bibinfo{author}{\bibfnamefont{I.}~\bibnamefont{Martin}},
  \bibinfo{journal}{Phys. Rev. B} \textbf{\bibinfo{volume}{73}},
  \bibinfo{pages}{035104} (\bibinfo{year}{2006}).

\bibitem[{\citenamefont{Ryndyk et~al.}(2008)\citenamefont{Ryndyk, D'Amico,
  Cuniberti, and Richter}}]{Ryndyk08prb}
\bibinfo{author}{\bibfnamefont{D.~A.} \bibnamefont{Ryndyk}},
  \bibinfo{author}{\bibfnamefont{P.}~\bibnamefont{D'Amico}},
  \bibinfo{author}{\bibfnamefont{G.}~\bibnamefont{Cuniberti}},
  \bibnamefont{and} \bibinfo{author}{\bibfnamefont{K.}~\bibnamefont{Richter}},
  \bibinfo{journal}{Phys. Rev. B} \textbf{\bibinfo{volume}{78}},
  \bibinfo{pages}{085409} (\bibinfo{year}{2008}).

\bibitem[{\citenamefont{D'Amico et~al.}(2008)\citenamefont{D'Amico, Ryndyk,
  Cuniberti, and Richter}}]{DAmico08njp}
\bibinfo{author}{\bibfnamefont{P.}~\bibnamefont{D'Amico}},
  \bibinfo{author}{\bibfnamefont{D.~A.} \bibnamefont{Ryndyk}},
  \bibinfo{author}{\bibfnamefont{G.}~\bibnamefont{Cuniberti}},
  \bibnamefont{and} \bibinfo{author}{\bibfnamefont{K.}~\bibnamefont{Richter}},
  \bibinfo{journal}{New J. Phys.} \textbf{\bibinfo{volume}{10}},
  \bibinfo{pages}{085002} (\bibinfo{year}{2008}).

\bibitem[{\citenamefont{Repp et~al.}(2004)\citenamefont{Repp, Meyer, Olsson,
  and Persson}}]{Repp04science}
\bibinfo{author}{\bibfnamefont{J.}~\bibnamefont{Repp}},
  \bibinfo{author}{\bibfnamefont{G.}~\bibnamefont{Meyer}},
  \bibinfo{author}{\bibfnamefont{F.~E.} \bibnamefont{Olsson}},
  \bibnamefont{and} \bibinfo{author}{\bibfnamefont{M.}~\bibnamefont{Persson}},
  \bibinfo{journal}{Science} \textbf{\bibinfo{volume}{305}},
  \bibinfo{pages}{493} (\bibinfo{year}{2004}).

\bibitem[{\citenamefont{Olsson et~al.}(2007)\citenamefont{Olsson, Paavilainen,
  Persson, Repp, and Meyer}}]{Olsson07prl}
\bibinfo{author}{\bibfnamefont{F.~E.} \bibnamefont{Olsson}},
  \bibinfo{author}{\bibfnamefont{S.}~\bibnamefont{Paavilainen}},
  \bibinfo{author}{\bibfnamefont{M.}~\bibnamefont{Persson}},
  \bibinfo{author}{\bibfnamefont{J.}~\bibnamefont{Repp}}, \bibnamefont{and}
  \bibinfo{author}{\bibfnamefont{G.}~\bibnamefont{Meyer}},
  \bibinfo{journal}{Phys. Rev. Lett.} \textbf{\bibinfo{volume}{98}},
  \bibinfo{eid}{176803} (\bibinfo{year}{2007}).

\bibitem[{\citenamefont{Lang and Firsov}(1963)}]{Lang63jetp}
\bibinfo{author}{\bibfnamefont{I.~G.} \bibnamefont{Lang}} \bibnamefont{and}
  \bibinfo{author}{\bibfnamefont{Y.~A.} \bibnamefont{Firsov}},
  \bibinfo{journal}{Sov. Phys. JETP} \textbf{\bibinfo{volume}{16}},
  \bibinfo{pages}{1301} (\bibinfo{year}{1963}).

\bibitem[{\citenamefont{Hewson and Newns}(1974)}]{Hewson74jjap}
\bibinfo{author}{\bibfnamefont{A.~C.} \bibnamefont{Hewson}} \bibnamefont{and}
  \bibinfo{author}{\bibfnamefont{D.~M.} \bibnamefont{Newns}},
  \bibinfo{journal}{Japan. J. Appl. Phys.} \textbf{\bibinfo{volume}{Suppl.\,2,
  Pt.\,2}}, \bibinfo{pages}{121} (\bibinfo{year}{1974}).

\bibitem[{\citenamefont{Mahan}(1990)}]{Mahan90book}
\bibinfo{author}{\bibfnamefont{G.}~\bibnamefont{Mahan}},
  \emph{\bibinfo{title}{Many-Particle Physics}} (\bibinfo{publisher}{Plenum,
  New York}, \bibinfo{year}{1990}), \bibinfo{edition}{2nd} ed.

\bibitem[{\citenamefont{Braig and Flensberg}(2003)}]{Braig03prb}
\bibinfo{author}{\bibfnamefont{S.}~\bibnamefont{Braig}} \bibnamefont{and}
  \bibinfo{author}{\bibfnamefont{K.}~\bibnamefont{Flensberg}},
  \bibinfo{journal}{Phys. Rev. B} \textbf{\bibinfo{volume}{68}},
  \bibinfo{pages}{205324} (\bibinfo{year}{2003}).

\bibitem[{\citenamefont{Mitra et~al.}(2004)\citenamefont{Mitra, Aleiner, and
  Millis}}]{Mitra04prb}
\bibinfo{author}{\bibfnamefont{A.}~\bibnamefont{Mitra}},
  \bibinfo{author}{\bibfnamefont{I.}~\bibnamefont{Aleiner}}, \bibnamefont{and}
  \bibinfo{author}{\bibfnamefont{A.~J.} \bibnamefont{Millis}},
  \bibinfo{journal}{Phys. Rev. B} \textbf{\bibinfo{volume}{69}},
  \bibinfo{pages}{245302} (\bibinfo{year}{2004}).

\bibitem[{\citenamefont{Koch and von Oppen}(2005)}]{Koch05prl}
\bibinfo{author}{\bibfnamefont{J.}~\bibnamefont{Koch}} \bibnamefont{and}
  \bibinfo{author}{\bibfnamefont{F.}~\bibnamefont{von Oppen}},
  \bibinfo{journal}{Phys. Rev. Lett.} \textbf{\bibinfo{volume}{94}},
  \bibinfo{pages}{206804} (\bibinfo{year}{2005}).

\bibitem[{\citenamefont{Koch et~al.}(2006{\natexlab{a}})\citenamefont{Koch,
  Semmelhack, von Oppen, and Nitzan}}]{Koch06prb}
\bibinfo{author}{\bibfnamefont{J.}~\bibnamefont{Koch}},
  \bibinfo{author}{\bibfnamefont{M.}~\bibnamefont{Semmelhack}},
  \bibinfo{author}{\bibfnamefont{F.}~\bibnamefont{von Oppen}},
  \bibnamefont{and} \bibinfo{author}{\bibfnamefont{A.}~\bibnamefont{Nitzan}},
  \bibinfo{journal}{Phys. Rev. B} \textbf{\bibinfo{volume}{73}},
  \bibinfo{pages}{155306} (\bibinfo{year}{2006}{\natexlab{a}}).

\bibitem[{\citenamefont{Koch et~al.}(2006{\natexlab{b}})\citenamefont{Koch, von
  Oppen, and Andreev}}]{Koch06prb2}
\bibinfo{author}{\bibfnamefont{J.}~\bibnamefont{Koch}},
  \bibinfo{author}{\bibfnamefont{F.}~\bibnamefont{von Oppen}},
  \bibnamefont{and} \bibinfo{author}{\bibfnamefont{A.~V.}
  \bibnamefont{Andreev}}, \bibinfo{journal}{Phys. Rev. B}
  \textbf{\bibinfo{volume}{74}}, \bibinfo{pages}{205438}
  (\bibinfo{year}{2006}{\natexlab{b}}).

\end{thebibliography}

\end{document}